# X-ray Spectroscopy of a Rare-Earth Molecular System Measured at the Single Atom Limit in Room Temperature

Sarah Wieghold[1,*], Nozomi Shirato[2], Xinyue Cheng[3], Kyaw Zin Latt[2], Daniel Trainer[2], Richard Sottie[4], Daniel Rosenmann[2], Eric Masson[3], Volker Rose[1] & Saw Wai Hla[2,4,*]

[1]Advanced Photon Source, Argonne National Laboratory, Lemont, IL 60439, USA

[2]Nanoscience & Technology Division, Argonne National Laboratory, Lemont, IL 60439, USA

[3]Department of Chemistry and Biochemistry, Ohio University, Athens, OH 45701, USA

[4]Nanoscale & Quantum Phenomena Institute, and Department of Physics & Astronomy, Ohio University, Athens, OH 45701, USA

*Corresponding author emails: swieghold@anl.gov, hla@ohio.edu

ABSTRACT

We investigate the limit of X-ray detection at room temperature on rare-earth molecular films using lanthanum and a pyridine-based dicarboxamide organic linker as a model system. Synchrotron X-ray scanning tunneling microscopy is used to probe the molecules with different coverages on a HOPG substrate. X-ray-induced photocurrent intensities are measured as a function of molecular coverage on the sample allowing a correlation of the amount of La ions with the photocurrent signal strength. X-ray absorption spectroscopy shows cogent $M_{4,5}$ absorption edges of the lanthanum ion originated by the transitions from the $3d_{3/2}$ and $3d_{5/2}$ to 4f orbitals. X-ray absorption spectra measured in the tunneling regime further reveal an X-ray excited tunneling current produced at the $M_{4,5}$ absorption edge of La ion down to the ultimate atomic limit at room temperature.

**INTRODUCTION**

Rare-earth metals exhibit promising optical, magnetic, and catalytic properties and are used in a wide variety of energy conversion applications and quantum technologies including photon upconversion,[1–3] quantum cutting,[4,5] memory storage,[6,7] and quantum networks.[8,9] Rare-earth atoms doped or implanted as 'defects' into a host crystal were demonstrated to exhibit long spin coherence and lifetimes,[10,11] narrow optical transitions, and the Laporte forbidden transition[12,13] owing to their unique electronic configuration of the 4f-shell electrons. Despite these unique properties, one drawback is related to the fabrication of defect centers with spatial control and selectivity. Recently, molecular systems such as coordination complexes, metal-organic frameworks, and supramolecular networks have attracted attention due to their potential to precisely control the location and local environment of the rare-earth atoms by incorporating them into molecular scaffolds.[14-16] These molecular approaches provide tantalizing engineering control in which the properties and functionalities of rare-earth metal ions can be tailored for specific applications.[17] For example, helicates, cages, or cubes have been used as building blocks for extended and well-designed molecular systems by employing tridentate ligands based on pyridine-diamide and -dicarbonyl or carboxylic acid chelating moieties.[18–20] Rare-earth-based molecular systems can also be used to control the distance between electron and nuclear spin qubits by changing the length of the organic linkers, allowing the rate and efficiency of coherence transfer between them to be tuned.[17] By introducing functionalized groups via C≡C or C≡N bridges to the linker moiety embedded into the molecular systems, even vibrational modes can be added to modulate the coherence time.[17,21] Additionally, due to the multimodality of the linkers based on their design, a variety of lanthanide (Ln) atoms can be placed with atomic precision into molecular systems enabling access to various possible spin states as well as providing synthetic opportunities for potential applications in emission, energy up-conversion, spintronic and quantum devices.

Here, we develop lanthanum and a pyridine-based dicarboxamide organic linker (L) as a model system, LnL$_3$, to investigate the limit of X-ray detection on rare-earth containing molecules at room temperature using synchrotron X-ray scanning tunneling microscopy (SX-STM). La is the lanthanide series' first element with an empty 4f shell with a [Xe]5d$^1$6s$^2$ configuration. The 4f shell can become partially occupied due to the hybridization of the 6s, 5d, and 4f electrons. The local environments of the rare-earth ions in the molecules are important to investigate for designing molecule-based functional devices containing rare-earth metals. Synchrotron X-rays offer a powerful experimental method to simultaneously investigate elemental, chemical, and magnetic properties of materials.[22-27] Recently, the development of the SX-STM technique opened further possibilities for the X-ray characterization of materials down to the atomic limit at a low temperature of ~30 K.[22] Here, we demonstrate that X-ray spectroscopy at an atomic limit can be performed at room temperature. This work also provides a fundamental understanding of the local chemical environment of rare-earth ions coordinated with the molecules and will be useful for designing novel rare-earth molecular systems.[28,29]

**EXPERIMENTAL METHODS**

To check the structure of the molecules, La(pcam)$_3$-(CF$_3$SO$_3$)$_3$ salt was deposited from a custom-built Knudsen cell under an ultrahigh vacuum (UHV) environment onto an atomically cleaned Cu(111) substrate. The sample was then transferred in situ to the scanning tunneling microscope (STM) chamber directly attached to the sample preparation chamber via a gate valve. STM imaging was performed at 5 K substrate temperature in UHV using a Createc GmbH STM system.

For the SX-STM measurements, La(pcam)$_3$ was dissolved in ethanol and the solution was then drop-casted onto HOPG. SX-STM measurements were performed at the XTIP beamline located at sector 4-ID-E of the Advanced Photon Source and the Center for Nanoscale Materials at the Argonne National Laboratory.[30] The SX-STM is operated in UHV at room temperature. X-ray absorption spectra (XAS) were recorded with a step size of 0.1 eV at a resolving power of E/ΔE of 4000.[23] For the experiments in the *near-field*, i.e., tunneling regime, the STM tip was positioned at a fixed tip height above the sample at a tunneling distance (~ 0.5 nm) using a tunneling current setpoint of 100 pA and a bias voltage of -1 V for data collection. Atomic force microscopy (AFM) measurements were performed in ScanAsyst mode using a Si cantilever (Veeco Multimode SPM, Bruker) at room temperature in ambient conditions.

**RESULTS AND DISCUSSION**

The coordination complex used for the model study here is La(pcam)$_3$ [Lanthanum (III) tris(2,6-pyridinedicarboxamide)]. La(pcam)$_3$ was synthesized from a solution of pcam and Lanthanum$^{III}$ trifluoromethane sulfonate in acetonitrile, which was concentrated under reduced pressure (Supplementary Information). La(pcam)$_3$ has three equivalent ligand arms in a planar, distorted D$_{3h}$ geometry with 120-degree angles between the nearest arms (Fig. 1a). The La ion is well protected by the ligands (Fig. 1a, and 1b). The structure of the synthesized molecules was initially investigated by STM imaging. For this part of the study, a very low coverage of molecules was deposited by thermal evaporation onto an atomically clean Cu(111) surface in a UHV environment. The STM images acquired at a substrate temperature of 5 K show scattered molecules on the surface (Fig. 1c). A closeup STM image (Fig. 1d) reveals the expected triangular shape of the molecule.

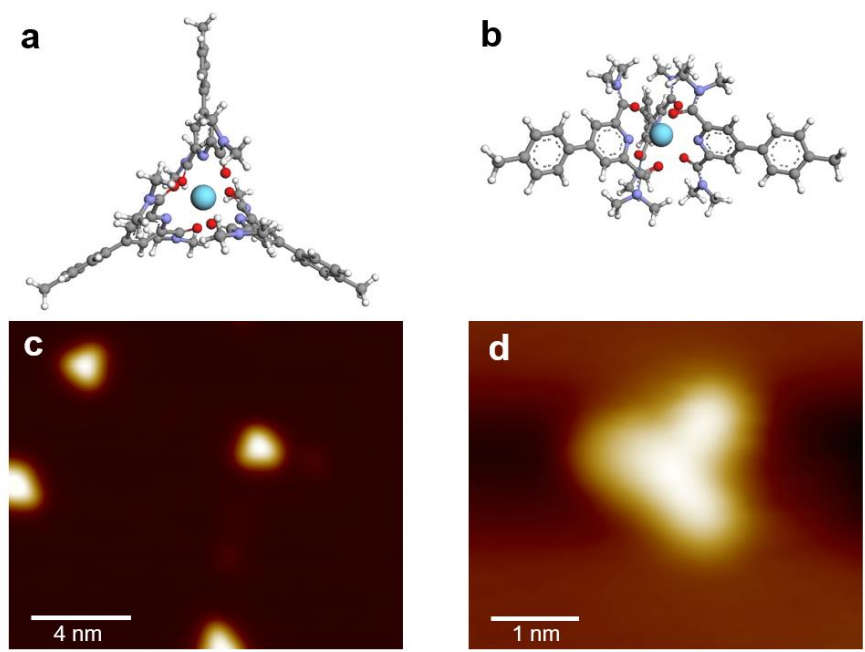

*Figure 1.* Structure of La(pcam)$_3$. Models showing *(a)* top view and *(b)* side view of La(pcam)$_3$. *(c)* An STM image displaying scattered La(pcam)$_3$ molecules on a Cu(111) surface at 5 K. *(d)* A zoomed-in STM image shows an isolated La(pcam)$_3$ on Cu(111).

After successful synthesis and structural characterization of the La(pcam)$_3$ molecular system, we moved on to perform molecular coverage-dependent SX-STM measurements. For the experiments, different La(pcam)$_3$ coverages were drop-casted from the solution onto a freshly cleaved highly ordered pyrolytic graphite (HOPG) substrate. For calibration, we first drop-cast solvent only onto the HOPG substrate and checked the surface with AFM imaging at ambient conditions. Large area AFM images show no apparent solvent features, and the step-edges of the surface can be clearly observed (Fig. 2a). After drop-casting solvent containing La(pcam)$_3$, the AFM images reveal relatively flat but brighter protruding regions partially covered across the surface (Fig. 2b). The La(pcam)$_3$ coverage was estimated from large area AFM images (Fig. 2 and Supplementary Information).

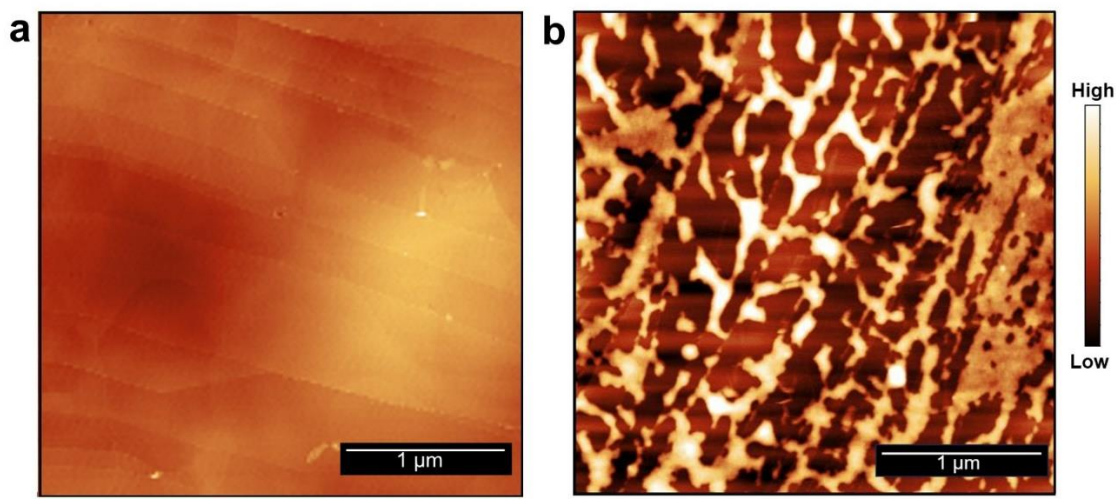

*Figure 2. AFM characterizations. (a) AFM images of HOPG acquired after solvent-only deposition shows step-edge features of the substrate. (b) After drop-casting La(pcam)$_3$, the areas covered by the molecules appear as protrusions (light color areas).*

For our SX-STM experiments, a monochromatic synchrotron soft X-ray beam with a size of ~10 µm x 10 µm cross-section is passed through a high-frequency X-ray chopper and then it is focused onto the tip-sample junction with an approximate angle of 12.5 degrees from the surface plane. The X-ray chopper generates on-and-off cycles of the X-ray beam. When the X-ray beam is in the "off" position, an STM feedback loop is active for maintaining the tip height in the tunneling regime. X-ray excited tunneling electrons are collected using a lock-in amplifier when the shutter is in the "on" position. The frequency of the chopper is synchronized with the lock-in amplifier and only the X-ray excited current is collected by rejecting the background current. X-ray absorption spectra are obtained by sweeping the photon energy over the La M$_{4,5}$ edge region, from 832.5 eV to 857.5 eV.

In our SX-STM setup, the tip and sample photocurrents can be detected simultaneously (Fig. 3a) using lock-in amplifiers connected to the tip and sample channels separately.[22,31] This enables a comparative study between the X-ray excited current signals recorded at the tip and sample channels. Generally, two measurement modes of SX-STM, *far-field*, and *near-field*, can be realized.[22,31] In the *far-field* mode, the STM tip is held a few nanometers above the surface, out of the electron tunneling range (Fig. 3a). When incident X-ray photons produce photoelectrons with energies higher than the work function of the sample, these electrons escape from the sample. In the soft X-ray regime, a significant number of Auger and secondary electrons are also generated during this process, which already occurs when core-level electrons are excited to the states between the Fermi level and the work function.[32] The photoexcitation of a core-level electron to either unoccupied states between the Fermi level and the work function, or to the continuum, i.e., above the work function, leaves a core hole that is subsequently filled by electrons from the higher orbits. During this process, both Auger and secondary electrons are produced,[22, 32] and electrons with sufficient energy also escape the sample. A net current is generated by the electrons leaving the sample, and it is directly recorded by a lock-in amplifier at the sample channel (Fig. 3a) as a function of incident photon energy (Fig. 3b). The photocurrent signal recorded at the sample channel here is similar to the standard total electron yield (TEY) mode measurements.[32,33]

The sample channel measures the total number of electrons that are lost from the entire sample area illuminated by the X-ray beam, while the simultaneously recorded tip channel in the *far-field* mode captures only a fraction of these photo-ejected electrons.[22] Here, the utilization of a specialized coaxial tip enables the detection of a small fraction of photocurrent being captured by the tip (Fig. 3c). The tip used in this study is formed by a tungsten (W) core, which is coated with a $SiO_2$ insulating layer to prevent the collection of photo-ejected electrons at the side wall of the tip.[22,24] The outer wall of the coaxial tip is then coated with a titanium buffer layer and a gold layer, which is grounded to prevent capacitive charging. Only a range of 30 to 200 nm of the W tip apex is exposed, and thus it enables collecting photoelectrons locally.

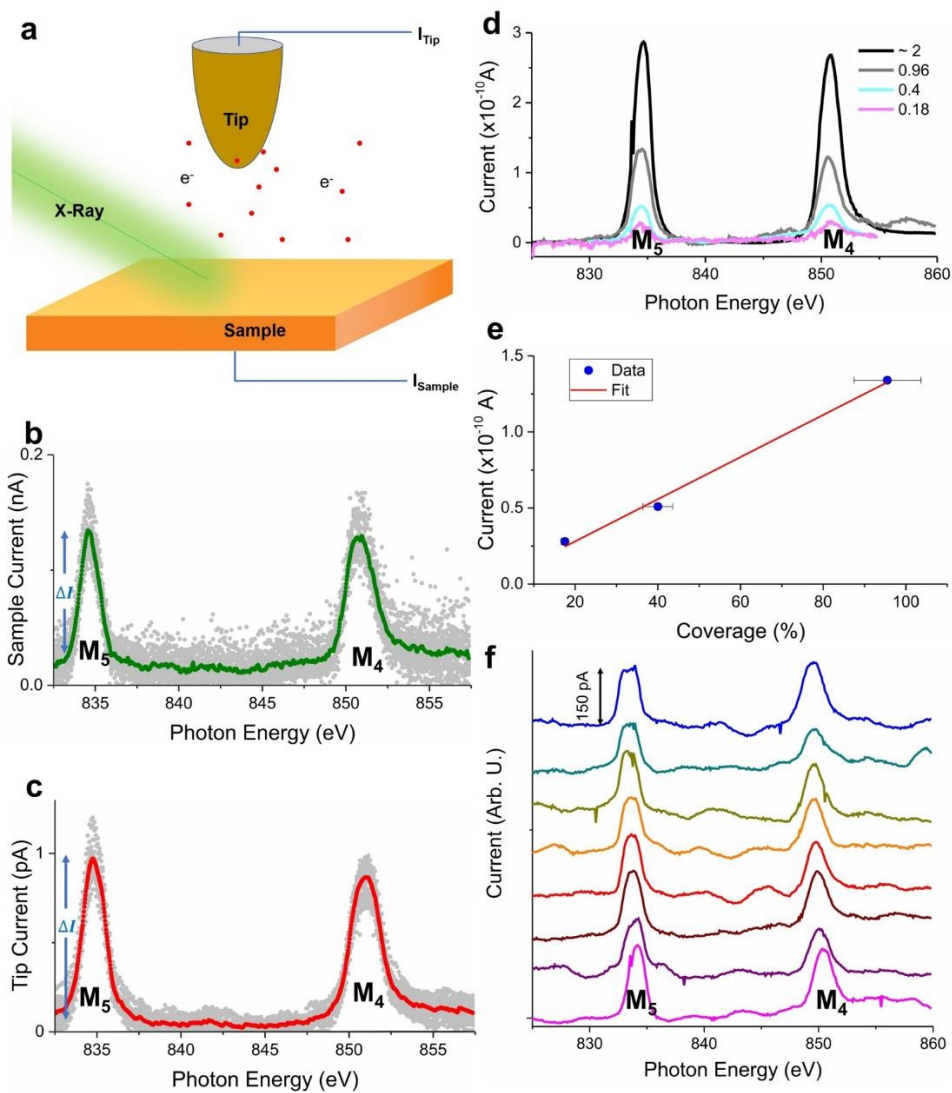

*Figure 3. STM-XAS in the far-field regime. (**a**) Demonstration of SX-STM set-up in the far-field regime where the tip is located ~5 nm above the surface (out of the tunneling range). XAS spectra of the La $M_{4,5}$ absorption edge were measured at the sample channel (**b**), and the tip channel (**c**). (**d**) XAS spectra for molecular coverages of ~2, 0.96, 0.4, and 0.18 on HOPG. Here, the initial current for each curve is set to zero. (**e**) Current vs. La(pcam)$_3$ coverage plot. (**f**) XAS spectra acquired at different locations on the sample locations show similar $M_{4,5}$ peak intensities. In (b), (c), and (d), the initial current is set to zero.*

Initially, we measure the $M_{4,5}$ absorption edge of La to probe the *d* to *f* transitions in TEY mode, i.e., recording the photocurrent at the sample channel only in the *far-field* mode (Fig. 3b). Due to spin-orbit coupling, the $M_5$ and $M_4$ edges originate from $3d_{5/2}$ and $3d_{3/2}$ transitions to *4f* orbitals, respectively.[22,34] The peaks corresponding to $M_5$ and $M_4$ edges can be clearly observed in both the sample and tip channels (Fig. 3b, and 3c). For 0.96 La(pcam)$_3$ coverage, the total current at the sample channel for the $M_5$ peak is measured as, $\Delta I_{sample}$= 134±31 pA (Fig. 3b), while the simultaneously recorded tip channel shows the maximum current for the $M_5$ peak of $\Delta I_{tip}$= 1.0±0.2 pA (Fig. 3c). Thus, the total number of photo-ejected electrons captured in the tip channel is about 2 order of magnitude less than the standard TEY mode of the sample channel.

Next, XAS spectra are acquired for various coverages of La(pcam)$_3$ on HOPG. Figure 3d shows the XAS spectra for La(pcam)$_3$ coverages of 2, 0.96, 0.4, and 0.17 monolayers. Here, the coverage of the fully covered molecular layer is estimated as ~ 2 monolayers. As expected, the maximum current at the $M_5$ peak is progressively lower as the coverage decreases. Figure 3e shows the plot of the sample current measured at the $M_5$ peak as a function of sub-monolayer coverages of 0.96, 0.4, and 0.17. The coverage of 2 is excluded in the plot, because of the uncertainty in the estimation of the full layer coverage. Obviously, the current linearly increases with the La(pcam)$_3$ coverage. XAS spectra are also measured at different locations on the sample, which reveal similar $M_{4,5}$ absorption edge intensities (Fig. 3f). This indicates the uniformity of La(pcam)$_3$ molecules concentration across the sample after drop-casting.

Recently, X-ray spectroscopy measurements on an iron and a terbium atom using SX-STM technique at low temperature (~30 K) have been demonstrated.[22] The next step in this research direction is to answer whether similar single-atom X-ray detection can be realized at room temperature. For this measurement, the SX-STM is operated in the *near-field*, i.e., in the tunneling regime (Fig. 4a) at room temperature. Here, the SX-STM tip is approached to the sample to a tunneling distance, ~ 0.5 nm, using -1 V bias and 100 pA tunneling current. Then the images of the sample area to be examined are acquired in STM imaging mode. The images (Fig. 4b) reveal that the La(pcam)$_3$ molecules form a disordered layer on HOPG after drop-casting. Because of its disordered nature, it is difficult to resolve individual molecules although the shape of the molecules can be identified occasionally (Fig. 4b). This can be confirmed by comparison with the single-molecule images shown in Fig. 1. For a conventional STM analysis, such a sample condition is challenging, because the molecular layer may also include trapped solvent molecules. However, the SX-STM probes core-level electrons. Thus, only the La ions in the layer provide the desired signal at the characteristic photon energy. Thus, unlike STM, SX-STM is greatly advantageous in the characterization of materials.

Next, the X-rays beam illuminates the tip-sample junction of the 0.96 coverage sample while the tip remains static above the molecular layer at a tunneling distance. Then the X-ray beam energy is ramped from 832.5 eV to 857.5 eV to cover the La $M_{4,5}$ absorption edges. The X-ray excited tunneling electrons are simultaneously recorded at the tip and the sample channels using separate lock-in amplifiers as before. For reproducibility, the STM-XAS spectra are collected every 6 nm across the sample for a total length of 150 nm. The results reveal two different sets of data (Fig. 4c to 4f).

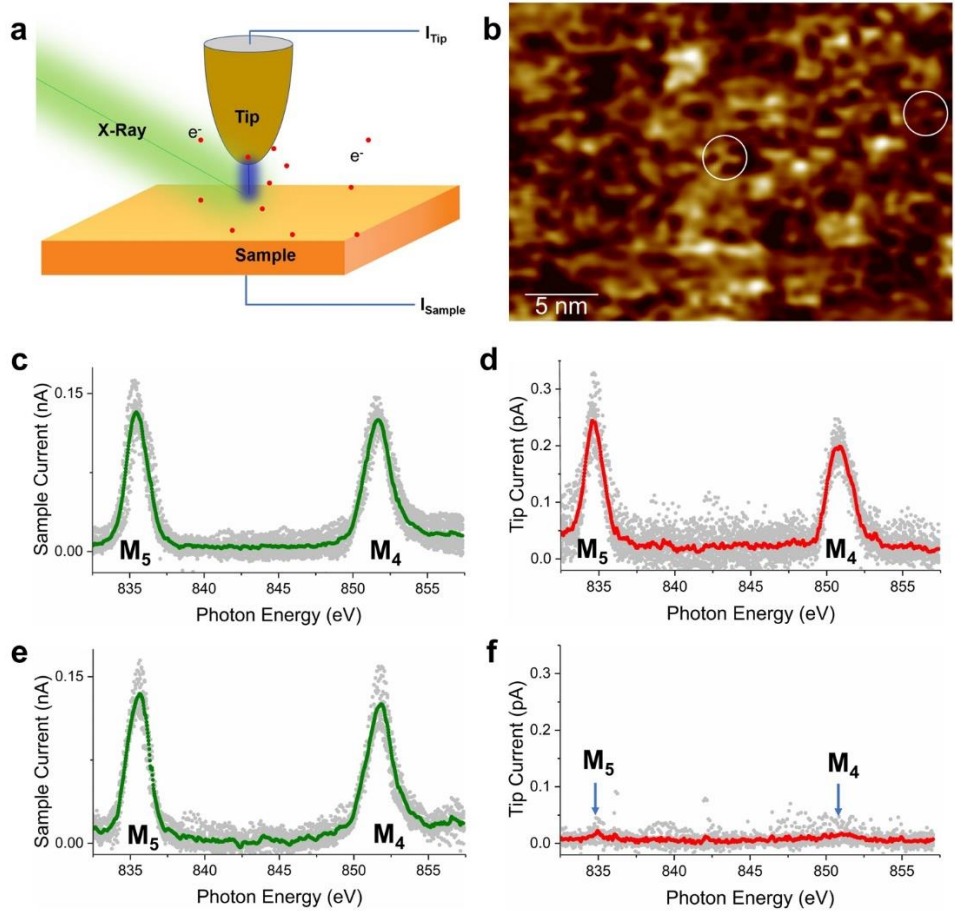

***Figure 4.*** *STM-XAS in the tunneling regime. (**a**) Demonstration of SX-STM set-up in near-field (tunneling) regime. (**b**) STM image of La(pcam)$_3$ layer (0.96 coverage) drop-casted on HOPG measured by SX-STM. (**c**) STM-XAS spectra of La(pcam)$_3$ were measured at the sample channel when the tip signals show the M$_{4,5}$ edges. (**d**) STM-XAS spectra recorded at the tip channel showing cogent M$_{4,5}$ edges in the tunneling range. (**e**) STM-XAS spectra of La(pcam)$_3$ were measured at the sample channel when the tip signals do not show the M$_{4,5}$ edges. (**f**) STM-XAS spectra recorded at the tip channel do not show cogent M$_{4,5}$ edges in the tunneling range. The STM coaxial tip is held in the tunneling range, $I_t = 100$ pA, $V_t = -1$ V. The green curves in (c) and (e), and the red curves in (d) and (f) are the average data. In (c, d, e, and f), the initial current is set to zero.*

In the first set of spectra, both the sample and tip channels show cogent M$_{4,5}$ edge signals of La (Fig. 4c and 4d). Here, the sample channel in the tunneling regime contains contributions from both photo-ejected electrons and X-ray-excited tunneling electrons. It can be expressed as,

$$I_{\text{total}} = I_{\text{sample}} + I_{\text{x-tunnel}} \quad \text{----(1)}$$

Here, $I_{sample}$ is the sample current purely composed of X-ray ejected electrons, which is produced by the entire X-ray illuminated area and it is similar to the TEY mode, while $I_{x-tunnel}$ is the X-ray excited tunneling current that is atomically localized. Note that the X-rays also excite electrons from the tip however the resultant current produced by the electrons leaving from the tip does not produce the La edge signal and thus it enters as the background, which is subtracted.

From Fig. 3b, an $I_{sample}$ value of 0.134±0.031 nA is measured, while $I_{x-tunnel}$ = 0.23±0.07 pA is directly determined from Fig. 4d. Because $I_{sample}$ is about 3 order of magnitude higher than $I_{x-tunnel}$, $I_{total} \approx I_{sample}$. Thus, the recorded STM-XAS spectra for the sample channels in tunneling regime (Fig. 4c and 4e) appear with a similar intensity as in Fig. 3b. The second set of spectra (Fig. 4e, and 4f) reveal the $M_{4,5}$ edge signals of La only at the sample channel but there is no clear edge signal in the tip spectra although small traces of the edge signals can be discerned (indicated with arrows in Fig. 4f).

Quantum tunneling is sensitive to the atomic position of the surface underneath the tip. It is known that the tunneling current exponentially decays with distance, thus a change in 1 Å of tip height reduces the current approximately by one order of magnitude. Consequently, the top surface layer overwhelmingly contributes to the X-ray excited tunneling current while the contributions from the subsurface layers are negligible.[24] By considering the area of a single La(pcam)$_3$ molecule, 1.73 nm$^2$, where only one La atom is present, the contribution to the tunneling current by the La atoms from nearby molecules should be negligible. Thus, when the tip is directly located above a La ion in tunneling distance, the La edge signal can be obtained by the X-ray excited tunneling process while it is not the case when the tip is slightly displaced from the La ion position. Such evidence has been recently demonstrated for low-temperature SX-STM measurements of isolated Fe and Tb atoms coordinated to molecular hosts where the Fe and Tb edge signals can be observed only when the tip is directly located above the atomic positions in a tunneling distance, otherwise, their edge signals are absent.[22]

As discussed above, the far-field sample current is similar to a TEY mode measurement, and the current intensity depends on the amount of La atoms in the molecular film. The utilization of a specialized coaxial tip reduces about 2 orders of magnitude in the tip current in the *far-field* as compared to the sample channel. However, in the *near-field* regime, the current at the tip channel is produced by the X-ray excited tunneling process and it is inherently limited by the available states of a single atom responsible for the tunneling process. Thus, unlike the tip current intensity in the *far-field*, the tip current in the *near-field* is independent of the molecular coverage. The current at the $M_5$ edge in the tip channel in the *near-field* is 0.23±0.07 pA (Fig. 4d). This value agrees with the recently reported X-ray excited current values of individual Fe and Tb atoms, 0.3 pA and 0.1 to 0.2 pA, at their $L_3$ and $M_5$ edges, respectively.[22] This further confirms that the observed signal is produced from a single La ion.

**CONCLUSIONS**

We used the SX-STM technique to characterize a rare-earth-based molecular system at room temperature. By using various sub-monolayer coverages of La(pcam)$_3$ adsorbed on HOPG, we establish a linear relationship between the coverage and the overall signal strength in XAS spectra for a rare-earth element. Moreover, by extending measurements with SX-STM from the *far-field* to the *near-field* regime, we show that X-ray characterization can be performed down to a single

La ion via the X-ray excited tunneling process at room temperature. This work provides a fundamental understanding of the local environment of rare-earth ions caged inside molecular scaffolds and will pave the way for investigating novel rare-earth molecular systems.

**SUPPORTING INFORMATION DESCRIPTION**

1. Generalities
2. Preparation of pcam (3) and [La(pcam)$_3$](CF$_3$SO$_3$)$_3$
3. Characterization of [La(pcam)$_3$](CF$_3$SO$_3$)$_3$ (4) and precursors
4. Atomic force microscopy characterization of various coverages of La(pcam)$_3$ drop-casted onto HOPG
5. References for supporting information

**ACKNOWLEDGEMENTS**

We acknowledge financial support from the U.S. Department of Energy, Office of Science, Office of Basic Energy Sciences, Materials Science and Engineering Division. Work performed at the Center for Nanoscale Materials and Advanced Photon Source, both U.S. Department of Energy Office of Science User Facilities, was supported by the U.S. DOE, Office of Basic Energy Sciences, under Contract No. DE-AC02-06CH11357.

**TABLE OF CONTENTS IMAGE**

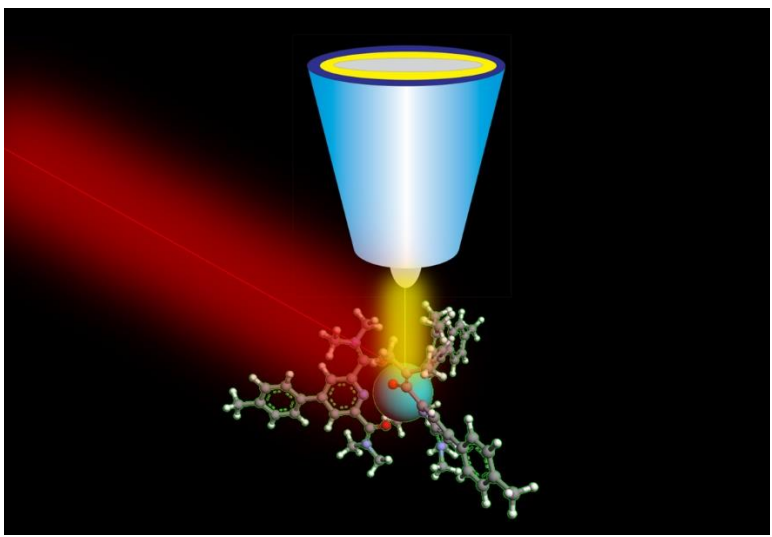